\documentclass[aps,10pt,prd,notitlepage,showpacs,nofootinbib,superscriptaddress]{revtex4-1}

\usepackage{graphicx}
\usepackage[utf8]{inputenc} 
\usepackage{amsmath}
\usepackage{amssymb}
\usepackage{float}
\usepackage{comment}
\usepackage{slashed}
\usepackage[normalem]{ulem}
\usepackage{cancel}
\makeatletter 
\renewcommand{\fnum@figure}{\textbf{Fig.~\thefigure}}
\makeatother
\usepackage{multirow}
\usepackage{rotating}
\usepackage[usenames,dvipsnames]{color}
\usepackage[colorinlistoftodos]{todonotes}
\usepackage[colorlinks=true,citecolor=blue,urlcolor=blue, pdfborder={0 0 0}]{hyperref}
\usepackage[normalem]{ulem}
\definecolor{darkred}{rgb}{0.6,0,0}
\usepackage[colorinlistoftodos]{todonotes}
\definecolor{linkcolor}{rgb}{0,0,0.5}

\usepackage{soul}
\definecolor{mightnightblue}{RGB}{25,25,112}

\definecolor{brown}{rgb}{0.59, 0.29, 0.0}

\begin{document}

\title{\color{BrickRed}CP-Violating and Charged Current Neutrino Non-standard Interactions in CE$\nu$NS}
\author{Amir N.\ Khan}
\email{amir.khan@mpi-hd.mpg.de}
\affiliation{Max-Planck-Institut f\"ur Kernphysik, Postfach 103980, D-69029
Heidelberg, Germany}
\author{Douglas W.\ McKay}
\email{dmckay@ku.edu}
\affiliation{Department of Physics and Astronomy, University of Kansas, Lawrence, KS
66045, USA}
\author{Werner Rodejohann}
\email {werner.rodejohann@mpi-hd.mpg.de}
\affiliation{Max-Planck-Institut f\"ur Kernphysik, Postfach 103980, D-69029
Heidelberg, Germany}

\begin{abstract}\noindent
Neutrino non-standard interactions (NSI) can be constrained using coherent elastic neutrino-nucleus scattering. We discuss here two  aspects in this respect, namely effects of (i) charged current NSI in neutrino production and (ii) CP-violating phases associated with neutral current NSI in neutrino detection. Effects of CP-phases require the simultaneous presence of two different flavor-changing neutral current NSI parameters. Applying these two scenarios to the COHERENT measurement, we derive limits on charged current NSI and find that more data is required to compete with the existing limits. Regarding CP-phases, we show how the limits on the NSI parameters depend dramatically on the values of the phases.Accidentally, the same parameters influencing coherent scattering also show up in neutrino oscillation experiments. We find that COHERENT provides complementary constraints on the set of NSI parameters that can explain the discrepancy in the best-fit value of the standard CP-phase obtained by T2K and NO$\nu$A, while the significance with which the LMA-Dark solution is ruled out can be weakened by the presence of additional NSI parameters introduced here. 
\end{abstract}
\date{\today }
\pacs{xxxxx}
\maketitle

\section{Introduction}
Coherent elastic neutrino-nucleon scattering (CE$\nu$NS) is an allowed  standard model (SM)  process which was predicted in the seventies 
\cite{Freedman:1973yd, Freedman:1977xn} and was observed very recently by the COHERENT experiment \cite{Akimov:2017ade, Akimov:2018ghi, Akimov:2018vzs}. In between the theoretical prediction and its observation, the formalism to use CE$\nu$NS as a probe for new neutrino physics, new neutral current physics or nuclear physics was pointed out for several scenarios \cite{Tubbs:1975jx, Drukier:1983gj, Barranco:2005yy, Scholberg:2005qs, Leitner:2006ww, Formaggio:2011jt, Anderson:2012pn, deNiverville:2015mwa,Kosmas:2015sqa, Dutta:2015nlo, Lindner:2016wff, Kosmas:2017zbh}. 
After its observation  there has been a surge of papers that study limits imposed by the COHERENT data on various standard and new physics aspects, see e.g.\  \cite{Dent:2016wcr,Coloma:2017ncl,Coloma:2017egw, AristizabalSierra:2017joc,Kosmas:2017tsq,Ge:2017mcq,Liao:2017uzy,Denton:2018xmq,Farzan:2018gtr,Abdullah:2018ykz, Billard:2018jnl,Farzan:2018gtr,Esteban:2018ppq,AristizabalSierra:2018eqm, Denton:2018xmq,Brdar:2018qqj,Billard:2018jnl,Gonzalez-Garcia:2018dep,Altmannshofer:2018xyo,Cadeddu:2018dux,Heeck:2018nzc,Khan:2019cvi,Cadeddu:2019qmv,Arcadi:2019uif,Alikhanov:2019drg,Bischer:2019ttk,Papoulias:2019xaw, Dutta:2019eml,AristizabalSierra:2019ykk,Giunti:2019xpr,Canas:2019fjw,Coloma:2019mbs,Coloma:2019mbs,Giunti:2019xpr,Denton:2020hop,Flores:2020lji,Miranda:2020zji,Tomalak:2020zfh,Skiba:2020msb,Suliga:2020jfa,Cadeddu:2020nbr,Coloma:2020gfv,EstevesChaves:2021jct,Shoemaker:2021hvm}.

In particular, non-standard interactions (NSI) are a popular new physics scenario that can be constrained by CE$\nu$NS. NSI arise for instance via effective dimension-6 interactions of neutrinos with terrestrial matter. Possible effects during neutrino production, propagation and detection  have been an important feature of neutrino phenomenology as reviewed in refs.\ \cite{Davidson:2003ha,Ohlsson:2012kf,Farzan:2017xzy}. Many theories beyond the SM generate NSI at some level. If present, they can lead in current and future neutrino oscillation experiments to modified or even wrong measurements of neutrino parameters \cite{Bergmann:1998ft,Johnson:1999ci,GonzalezGarcia:2001mp,Kopp:2007ne,Khan:2013hva,Girardi:2014kca,Agarwalla:2014bsa,deGouvea:2015ndi,Deepthi:2016erc,Coloma:2016gei,Bakhti:2016gic,Masud:2016nuj,Ghosh:2017lim,Capozzi:2019iqn,Dutta:2020che,Esteban:2020itz,Chatterjee:2020kkm,Esteban:2020itz}. 
In particular, NSI include additional CP-phases beyond the single phase relevant in the standard neutrino picture. In this respect it should be noted that a tension in the determination of the standard CP-phase in the T2K and NO$\nu$A experiments \cite{himmel:a,dunne:p} can be explained by neutral current NSI including a new CP-phase \cite{Denton:2020uda,Chatterjee:2020kkm}. 
Another feature concerns LMA-Dark, i.e.\ 
the octant of the "solar neutrino angle" $\theta_{12}$, which in the presence of flavor diagonal NSI  can be different ($\theta_{12}>\pi/4$) from the one in the standard picture ($\theta_{12}<\pi/4$)  \cite{Miranda:2004nb}.  In general, the degeneracies between standard and new parameters in neutrino oscillation probabilities need to be broken by complementary measurements, in particular by scattering experiments. Indeed, CE$\nu$NS may be crucial here, already providing limits that disfavor the LMA-Dark solution \cite{Coloma:2017egw,Coloma:2017ncl,Denton:2018xmq,Coloma:2019mbs,EstevesChaves:2021jct}. 

We will discuss in this paper two aspects of NSI in coherent scattering.  These are (i) effects of charged current NSI in the production of neutrinos, and (ii) effects of CP-phases of neutral current NSI in the detection of neutrinos.  To the best of our knowledge, charged current NSI were not studied in the context of CE$\nu$NS, and a dedicated paper of CP-phases associated with effective NC NSI does not exist either. Aspects of CP violation in coherent scattering were discussed, though, but in a slightly different context. In ref.\ \cite{AristizabalSierra:2019ufd} a light vector boson with complex couplings was considered, but no connection to oscillation physics was made. Ref.\ \cite{Denton:2020uda} mentions that the parameter values explaining the T2K/NO$\nu$A discrepancy can be tested in CE$\nu$NS, but does not study effects of the CP-phases in CE$\nu$NS. Finally, ref.\ \cite{Esteban:2019lfo} provides global fits of oscillation and COHERENT data with focus on CP violation, but fitted only the absolute values of the NSI parameters when using COHERENT data. 
Our goal here is to present a  formalism  which takes into account CC NSI in pion and muon decay at the spallation neutron source relevant for COHERENT, as well as NC NSI along with the new CP-phases for the detection process. We will confront this setup with the COHERENT data that used a 
CsI[Na] target \cite{Akimov:2017ade, Akimov:2018ghi, Akimov:2018vzs}.  
Limits are presented on CC NSI parameters. Effects of CP-phases from NC NSI require that at least two different flavor-changing NSI terms are present. We will demonstrate that in this case the constraints on the NSI parameters depend crucially on the values of the new CP-phases. We show as a further example that in this case COHERENT can set complementary limits to the  parameter space relevant for the T2K/NO$\nu$A discrepancy. Finally, we will  estimate how the exclusion level of LMA-Dark is reduced in case CC NSI and/or CP violating NC NSI are present.

This paper is organized as follows. In section \ref{sec:formalism} we begin by introducing the fitting procedure and develop the formalism to describe NC and CC at source and detector. In section \ref{sec:results} we discuss our results for CP violating NC NSI, and CC NSI, before summarizing in section \ref{sec:concl}.

\section{Formalism}\label{sec:formalism}
\subsection{Experimental details and fitting procedure}
In this section we provide details of the COHERENT data that we will fit, and on our  fitting procedure.
The COHERENT experiment measures coherent elastic neutrino-nucleus scattering. Neutrinos are provided from pions decaying at rest, which in turn are produced from the spallation neutron source. The data we will use in this paper was collected with a total number of $1.76\times 10^{23}$ of protons on target (pot) delivered to  liquid mercury \cite{Akimov:2017ade, Akimov:2018ghi, Akimov:2018vzs}.  Mono-energetic muon neutrinos $%
(\nu _{\mu })$ at $E_{\nu } = 29.8$~MeV are produced isotropically from pion decay 
at rest ($\pi ^{+}\rightarrow \mu ^{+}\nu _{\mu })$ followed by a delayed
isotropic flux of electron neutrinos ($\nu _{e})$ and muon anti-neutrinos ($\bar{\nu}%
_{\mu })$ produced subsequently by muon-decay at rest ($\mu ^{+}\rightarrow \nu
_{e}e^{+}\bar{\nu}_{\mu }$). All three flavors are intercepted by a 
CsI[Na] detector at a distance of $L=19.3$ m from the source\footnote{Recently new data was provided by COHERENT indicating at about $3\sigma$ a non-zero CE$\nu$NS cross section with argon \cite{Akimov:2020pdx}.}. 
For all practical purposes, the CsI will be considered as a target since the Na as a dopant contributes negligibly \cite{Akimov:2017ade}.
We do not consider the timing information between the prompt and delayed signal of our analysis, which is a small effect at the current precision level of COHERENT as noted e.g.\ in  \cite{Giunti:2019xpr}. The average production rate of the SNS neutrinos from the pion decay chain is $r=0.08$ neutrinos of each flavor per proton. 
The differential event rate, after taking into account the detection
efficiency $\epsilon (T)$, taken from Fig.~S9 in ref.~\cite{Akimov:2017ade}, of
COHERENT reads
\begin{equation}
\frac{dN_{\nu _{\alpha }}}{dT}=tN\int_{E_{\nu }^{\rm min }}^{E_{\nu }^{\rm max}}dE_{\nu }\frac{d\sigma }{dT}(E_{\nu },T)\frac{%
d\phi _{\nu _{\alpha }}(E_{\nu })}{dE_{\nu }}\epsilon (T),
\label{eq:eventrt}
\end{equation}%
where $d\sigma/dT(E_{\nu },T)$ is the differential cross section of CE$\nu$NS with respect to nuclear recoil, and  $d\phi _{\nu _{\alpha }}(E_{\nu })/dE_\nu$ is the flux with respect to neutrino energy.  
Further, $t=308.1$~days is the run time of the experiment, $N=\left(2m_{\mathrm{det}}/M_{\rm CsI}\right) N_{A}$ is the total number of target nucleons, $m_{\mathrm{det}}=14.57$ kg, $N_{A}$ is Avogadro's  number,  $M_{\rm CsI}$ is the molar mass of CsI, $E_{\nu }^{\min}= \sqrt{MT/2}$, $M$ is the mass of the target nucleus, $E_{\nu }^{\max }$ is the upper limit of the neutrino energy
which is 52.8 MeV for the delay signal and 29.8 MeV for the prompt signal. We take a recoiled energy window of 4 to 25 keV for the analysis. 

Our fitting procedure follows closely our earlier work \cite{Khan:2019cvi}. In particular, 
we apply here a recent  measurement from ref.\ \cite{Collar:2019ihs}, which includes energy-dependence of the quenching factor. The 
following relation between the nuclear recoil energy and the number of photo-electrons (p.e.) is used: 
\begin{equation}\label{qf}
n_{\rm p.e.} = f_{Q}(T)\times T \times \left(\frac{0.0134 }{\rm MeV}\right),
\end{equation}%
 where $ f_{Q}(T)$ is the new quenching factor and 0.0134 is the average yield of the scintillation light in the detector by a single electron per MeV; both values were taken from ref.\ \cite{Collar:2019ihs}. The expected number of events in the $i$-th bin, therefore, is
\begin{equation}
N^{\rm i} = \int_{T^{i}}^{T^{i+1}} \frac{dN_{\nu _{\alpha }}}{dT} \,dT, 
\label{eq:eventrtot}
\end{equation}%
where the nuclear recoil energy limits of the integration $(T^{i}, T^{i+1})$ for $i$-th bin are related to the corresponding limits in terms of number of photo-electrons by eq.\ (\ref{qf}). For the fitting analysis of the parameters we use the following $\chi ^{2}$-function  
\begin{equation}
\chi ^{2}=\underset{i=4}{\overset{20}{\sum }}\frac{%
[N_{\rm obs}^{i}-N_{\rm exp}^{i}(1+\alpha )-B^{i}(1+\beta )]^{2}}{(\sigma ^{i})^{2}}%
+\left( \frac{\alpha }{\sigma _{\alpha }}\right) ^{2}+\left( \frac{\beta }{%
\sigma _{\beta }}\right) ^{2}\,,  \label{eq:chisq}
\end{equation}%
where $N_{\rm obs}^{i}$ is the observed event rate in the $i$-th energy bin, 
$N_{\rm exp}^{i}$ is the expected event rate given  in eq.\ (\ref{eq:eventrt}) integrated over the recoiled energy corresponding to each flavor, and $B^{i}$  is the estimated 
background event number in the $i$-th energy bin extracted from Fig.\ S13 
of ref.\  \cite{Akimov:2017ade}. The statistical uncertainty in the $i$-th energy bin is $\sigma ^{i}$, 
and $\alpha $, $\beta $ are  pull parameters related to the signal systematic 
uncertainty and the background rates. The corresponding uncertainties of 
the pull parameters are $\sigma _{\alpha }=0.135$ \cite{Collar:2019ihs} and $\sigma _{\beta }=0.25$. 
We calculate $\sigma _{\alpha }$ by adding uncertainties related to flux (10\%), neutron capture (5\%), acceptance (5\%) and quenching factor (5.1\%) in quadrature. 

Having established the fitting procedure, we will now give the fluxes and the cross sections in the new physics scenarios that we are interested in, namely charged current non-standard interactions and neutral current non-standard interactions including new CP-phases. The former will modify the flux, $d\phi _{\nu _{\alpha }}(E_{\nu })/dE_\nu$, while the latter will modify the cross section, $d\sigma/dT(E_{\nu },T)$. 
\subsection{Effective Lagrangians and the NSI notations}
Neutrinos for the COHERENT setup originate from charged current (CC) reactions in pion ($\pi^+$) and muon ($\mu^+$) decays and are detected via neutral current (NC) interactions through coherent elastic scattering on the CsI[Na] target.  At the source, on top of the standard model weak interaction, there can be CC non-standard interactions (NSI) in the $\pi^+$ and $\mu^+$ decays. 
Those are described by effective dimension-6 terms \cite{Johnson:1999ci, Khan:2013hva, Khan:2014zwa,Khan:2016uon,Khan:2017oxw} as 

\begin{widetext}
\begin{eqnarray}
\mathcal{L}{\rm }^{\pi^+}_{\rm CC} &=&-\frac{G_{F}}{\sqrt{2}} \left(\delta_{\mu \beta}+\varepsilon _{\mu
\beta }^{udL}\right)\left[\bar{d}\gamma_{\lambda }(1-\gamma_{5})u\right]\left[\bar{\mu} \gamma^{\lambda }(1-\gamma_{5})\nu
_{\beta}\right],
\label{eq:PiLeff} 
\end{eqnarray}%
\end{widetext}

\begin{widetext}
\begin{eqnarray}
\mathcal{L}{\rm}^{\mu^+ }_{\rm CC} &=&-\frac{G_{F}}{\sqrt{2}} \left({\delta_{\alpha e}\delta_{\beta \mu}+\varepsilon _{\alpha \beta }^{\mu eL}}\right)\left[\bar{\nu}_{\alpha}\gamma _{\lambda }(1-\gamma_{5}) e\right]\left[\bar{\mu}\gamma ^{\lambda }(1-\gamma_{5})\nu _{\beta }\right].  
\label{eq:MuLeff} 
\end{eqnarray}%
\end{widetext}
Here $G_F$ is the Fermi constant, $\alpha, \beta$ denote the neutrino flavors ($e$, $\mu, \tau$),  and $\delta_{\alpha \beta}$ is the Kronecker delta. 
For example, in the presence of CC NSI the two body decay ($\pi ^{+}\rightarrow \mu^{+}\nu _{\mu })$ is modified to $\pi^{+}\rightarrow \mu ^{+}\nu _{\alpha }~(\alpha =e, \mu ,\tau )$, where  $\alpha =\mu$ corresponds to  a flavor-conserving NSI and $\alpha =e,\tau $ correspond to   flavor-changing NSI. In these three cases the parameters that control the fluxes are $\varepsilon _{\mu \mu }^{udL},\varepsilon _{\mu e }^{udL}$ and $\varepsilon _{\mu \tau }^{udL},$ respectively. Likewise, in the three-body leptonic decay of muons, the$\  \bar{\nu}_{\mu }$ flux is controlled by the parameters $\varepsilon _{\mu \mu }^{\mu e L},\  \varepsilon _{e\mu }^{\mu eL}$ and\ $\varepsilon_{\tau \mu }^{\mu e L}$, while the $\nu _{e}$ fluxes are controlled by $%
\varepsilon _{\mu e}^{\mu eL},\  \varepsilon _{ee}^{\mu eL}$ and\ $\varepsilon _{\tau e}^{\mu eL}$.

For the detection via NC reactions, non-standard interactions can modify it as well. 
At quark level, the NC NSI can be conveniently written as
\begin{widetext}
\begin{eqnarray}
\mathcal{L}{\rm}^{q}_{\rm NC} &=&-\frac{G_{F}}{\sqrt{2}} \left[\bar{\nu}_{\alpha}\gamma _{\lambda }(1-\gamma_{5}) {\nu_{\beta}}\right]\left[({\textit{\scalebox{1.1}g}_{L\alpha \beta} \delta_{\alpha \beta}+\varepsilon _{\alpha \beta }^{q L}})\bar{q}\gamma ^{\lambda }(1-\gamma_{5})q+({\textit{\scalebox{1.1}g}_{R\alpha \beta} \delta_{\alpha \beta}+\varepsilon _{\alpha \beta }^{q R}})\bar{q}\gamma ^{\lambda }(1+\gamma_{5})q\right].
\label{eq:NCLeff} 
\end{eqnarray}%
\end{widetext}
Here $q$ are first generation up/down quarks and \textit{${\scalebox{1.1}g}_{L/R\alpha \beta}$} are SM NC couplings with left/right-handed  target quarks.  
Indices $\alpha=\beta$ correspond to SM interactions plus flavor-conserving   NSI while $\alpha\neq\beta$ corresponds to pure beyond-the-standard-model  flavor-changing interactions. 
Summation over the flavor indices is implied in eqs.\ (\ref{eq:PiLeff}) - (\ref{eq:NCLeff}).  

 All $\varepsilon$ parameters are complex in the charged current interactions in eqs.\ (\ref{eq:PiLeff}) and (\ref{eq:MuLeff}). On the other hand, because of the hermiticity of the neutral current Lagrangian in eq. (\ref{eq:NCLeff}), all flavor-diagonal parameters are real while the flavor changing parameters are complex. Under the hermiticity condition, the latter interchange the flavor indices and the sign of the phases also changes, that is, particularly in eq.\  (\ref{eq:NCLeff}), 
$(\varepsilon _{\alpha \beta }^{q L/R})^\ast=\varepsilon _{\beta \alpha }^{qL/R}$ for $\alpha \neq \beta$. 

Often one rewrites the left- and right-handed $\varepsilon$ in vector and axial vector form. The effective interactions terms in eqs.\ (\ref{eq:PiLeff}) and (\ref{eq:NCLeff}) can be written as
\begin{widetext}
\begin{eqnarray}
\mathcal{L}{\rm }^{\pi^+}_{\rm CC} &=&-\frac{G_{F}}{\sqrt{2}} \left[\bar{\mu} \gamma^{\lambda }(1-\gamma_{5})\nu
_{\beta}\right]\left[(\delta_{\mu \beta}+\varepsilon _{\mu
\beta }^{udV})\bar{d}\gamma_{\lambda}u-(\delta_{\mu \beta}+\varepsilon _{\mu
\beta }^{udA})\bar{d}\gamma_{\lambda }\gamma_{5}u\right],
\label{eq:PiLeffva} 
\end{eqnarray}%
\end{widetext}
\begin{widetext}
\begin{eqnarray}
\mathcal{L}{\rm}^{q}_{\rm NC} &=&-\frac{G_{F}}{\sqrt{2}} \left[\bar{\nu}_{\alpha}\gamma _{\lambda }(1-\gamma_{5}) {\nu_{\beta}}\right]\left[({\textit{\scalebox{1.2}g}^{V}_{\alpha \beta} \delta_{\alpha \beta}+\varepsilon _{\alpha \beta }^{q V}})\bar{q}\gamma ^{\lambda }q+({\textit{\scalebox{1.2}g}^{A}_{\alpha \beta} \delta_{\alpha \beta}+\varepsilon _{\alpha \beta }^{q A}})\bar{q}\gamma ^{\lambda }\gamma_{5}q\right],
\label{eq:NCLeffVA} 
\end{eqnarray}%
\end{widetext}
where
\begin{widetext}
\begin{eqnarray}
\textit{\scalebox{1.1}g}^{V/A}_{\alpha \beta} \delta_{\alpha \beta}&=& \textit{\scalebox{1.1}g}^{L}_{\alpha \beta} \delta_{\alpha \beta} \pm \textit{\scalebox{1.1}g}^{R}_{\alpha \beta} \delta_{\alpha \beta} \,,
\label{gvaglr}
\end{eqnarray}%
\end{widetext}
and the vector and axial vector parameters are 
\begin{widetext}
\begin{eqnarray}
\varepsilon _{\alpha \beta }^{q V/A}&=& \varepsilon _{\alpha \beta }^{q L} \pm \varepsilon _{\alpha \beta }^{q R}.
\label{epsvaglr}
\end{eqnarray}%
\end{widetext}
We do not consider any right-handed currents in the pion decays, so the only remaining contribution is the left-handed one as given in eq.\ (\ref{eq:PiLeff}). On  top of this, since the pion is a pseudoscalar particle, only the axial vector part of the hadronic matrix element contributes in eq.\ (\ref{eq:PiLeffva}), and we also consider only the axial vector NSI. 
Likewise, for all practical purposes, the axial vector contribution in  CE$\nu$NS is negligibly small (see e.g.\ \cite{Lindner:2016wff}) and thus we will consider only the vector terms in eq.\ (\ref{eq:NCLeffVA}). That is,  we will consider for the CC NSI the parameters $\varepsilon_{\mu \beta }^{udA}$ and $\varepsilon_{\alpha \beta}^{\mu eL}$ for pion and muon decays at the neutrino production, while the NC NSI parameters are $\varepsilon_{\alpha \beta}^{qV}$  at the detection.  

\subsection{Fluxes with CC NSI, Cross Section with NC NSI and the Expected Energy Spectrum}

To estimate the effects of CC NSI at neutrino production, we have to include them in the charged current decays which will in turn modify the three fluxes in terms of the CC NSI parameters. There occur two types of parameters in each decay. One is a flavor diagonal  interaction which interferes with the standard model process, and the others are two flavor changing parameters for each decay. The contribution of the latter  adds incoherently to the SM.   
After adding both types of CC NSI effects in each decay, the total differential flux expression will change accordingly as
\begin{widetext}
\begin{eqnarray}\label{eq:fluxes}
\left[\frac{d\phi _{\nu_{\mu}}(E_{\nu })}{dE_{\nu }}\right]_{\rm NSI} &=&\left[\frac{d\phi _{\nu_{\mu}}(E_{\nu })}{dE_{\nu }}\right]_{\rm SM}\left[\left(\left \vert 1+\varepsilon _{\mu\mu }^{udA}\right \vert ^{2}+\left \vert \varepsilon _{\mu e }^{udA}\right \vert ^{2}+
\left \vert\varepsilon _{\mu\tau }^{udA}\right \vert ^{2}\right) \equiv 1+ 
2Re(\varepsilon_{\mu\mu}^{udA})+\underset{\alpha = e,\mu,\tau }{\sum }|\varepsilon_{\mu \alpha}^{udA}|^2\right], \notag\\ 
\left[\frac{d\phi _{\overline{\nu }_\mu }(E_{\nu })}{dE_{\nu }}\right]_{\rm NSI} &=&\left[\frac{d\phi _{\overline{\nu }_\mu }(E_{\nu })}{dE_{\nu }}\right]_{\rm SM} \left[\left(\left \vert 1+ \varepsilon _{\mu \mu}^{\mu eL}\right \vert ^{2}+ \left \vert \varepsilon _{\mu e}^{\mu eL}\right \vert ^{2}+ \left \vert \varepsilon _{\mu \tau}^{\mu eL}\right \vert ^{2}\right) \equiv 1+ 2Re(\varepsilon_{\mu\mu}^{\mu eL})+\underset{\alpha = e,\mu,\tau }{\sum }|\varepsilon_{\mu \alpha}^{\mu eL}|^2\right], \\ 
\left[\frac{d\phi _{\nu_e}(E_{\nu })}{dE_{\nu }}\right]_{\rm NSI} &=&\left[\frac{d\phi_{\nu_e}(E_{\nu })}{dE_{\nu }}\right]_{\rm SM} \left[\left(\left \vert 1+ \varepsilon _{e e}^{\mu eL}\right \vert ^{2}+\left \vert \varepsilon _{e\mu}^{\mu eL}\right \vert ^{2}+ \left \vert\varepsilon _{e\tau}^{\mu eL}\right \vert ^{2}\right) \equiv 1+
2Re(\varepsilon_{ee}^{\mu eL})+\underset{\alpha = e,\mu,\tau }{\sum }|\varepsilon_{e \alpha}^{\mu eL}|^2\right]\notag,
\end{eqnarray}
\end{widetext}
where the standard fluxes for COHERENT read 
\begin{widetext}
\begin{eqnarray}
\left[\frac{d\phi _{\nu_{\mu}}(E_{\nu })}{dE_{\nu }}\right]_{\rm SM} &=&\frac{rN_{\rm pot}}{4\pi L^{2}}%
\delta \left(E_{\nu }-\frac{m_{\pi }^{2}-m_{\mu }^{2}}{2m_{\pi }}\right),
\notag\\ 
\left[\frac{d\phi _{\overline{\nu }_{\mu}}(E_{\nu })}{dE_{\nu }}\right]_{\rm SM} &=&\frac{rN_{\rm pot}}{%
4\pi L^{2}}\frac{64E_{\nu }^{2}}{m_{\mu }^{3}}\left(\frac{3}{4}-\frac{E_{\nu}}
{m_{\mu}}\right),
\\
\left[\frac{d\phi_{\nu_{e}}(E_{\nu })}{dE_{\nu }}\right]_{\rm SM} &=&\frac{rN_{\rm pot}}{4\pi L^{2}}%
\frac{192E_{\nu }^{2}}{m_{\mu }^{3}}\left(\frac{1}{2}-\frac{E_{\nu }}{m_{\mu }}\right) \notag,
\label{eq:fluxessm}
\end{eqnarray}
\end{widetext}
with, again, $N_{\rm pot}=5.71\times 10^{20}$ being the number of protons per day, $%
L=19.3$ m is the baseline and $r=0.08$ is the number of neutrinos per flavor
per proton on target. In eq.\ (\ref{eq:fluxes}), for each flux there are only two types of parameters: twice the real part of the flavor diagonal NSI and the three modulus squared parameters which include one flavor diagonal and two flavor changing $\varepsilon$. 
Now we discuss the effect of NC NSI on the cross section of CE$\nu $NS. The differential cross section of CE$\nu $NS, with respect to the nuclear recoil energy $T$, for neutrinos with flavor $\beta$ and energy $E_{\nu}$ scattered off a
target nucleus $(A,Z)$, can be written for $T \ll M$ as~\cite{Freedman:1973yd, Freedman:1977xn, Tubbs:1975jx, Barranco:2005yy, Lindner:2016wff} 
\begin{equation}
\frac{d\sigma_{\beta}}{dT}(E_{\nu},T) \simeq \frac{G_{F}^{2}M}{\pi }Q_{W\beta
}^{2}\left( 1-\frac{MT}{2E_{\nu }^{2}}\right) F^{2}(q^{2})\,.
\label{eq:diff-crossec}
\end{equation}%
Here 
$M$ is mass of the target nucleus with  $Q_{W\beta}^{2}$ its weak nuclear charge, and $F(q^{2})$ is the nuclear form factor as a function of $q^2=2 M T$, the momentum transfer in the scattering of neutrinos off the nuclei. 
We take the nuclear form factor $F(q^{2})$ from ref.\ \cite{Klein:1999gv}, 
given by
\begin{equation}
F(q^{2})=\frac{4\pi \rho _{0}}{Aq^{3}}[\sin (qR_{A})-qR_{A}\cos (qR_{A})]%
\left[ \frac{1}{1+a^{2}q^{2}}\right] .  \label{F-bessel}
\end{equation}%
Here, $\rho _{0}$ is the normalized nuclear number density, $A$ is the  atomic number of CsI, $%
R_{A}=1.2A^{1/3}\, \mathrm{fm}$ is the nuclear radius, and $a=0.7\, \mathrm{fm
 }$ is the range of the Yukawa potential. 

The weak charge $Q_{W\beta}^{2}$ is expressed in terms of the proton number ($Z$), neutron number ($N$), standard vector coupling constants $%
g_{p}^{V}=1/2-2\sin ^{2}\theta _{W}$\footnote{We use the low
energy value $\sin^2\theta_W$ = 0.2387 \cite{Erler:2004in} for the analysis.}, $g_{n}^{V}=-1/2$ and the NC NSI parameters $\varepsilon _{\alpha \beta}^{uV}$ and $\varepsilon _{\alpha \beta}^{dV}$, as
\begin{equation}
Q_{W\beta }^{2} =\left[Z(g_{p}^{V}+2\varepsilon _{\beta \beta
}^{uV}+\varepsilon _{\beta \beta }^{dV})+
N(g_{n}^{V}+2\varepsilon _{\beta
\beta}^{dV}+\varepsilon _{\beta \beta }^{uV})\right]^{2} + \underset{\alpha \neq \beta }{\sum }
\left|Z(2\varepsilon _{\alpha \beta
}^{uV}+\varepsilon _{\alpha \beta }^{dV})+N(2\varepsilon _{\alpha
\beta}^{dV}+\varepsilon _{\alpha \beta }^{uV})\right|^{2}. 
\label{eq:Qall}
\end{equation}%
\begin{figure}[t]
\begin{center}
\includegraphics[width=6.0in, height=3.7in]{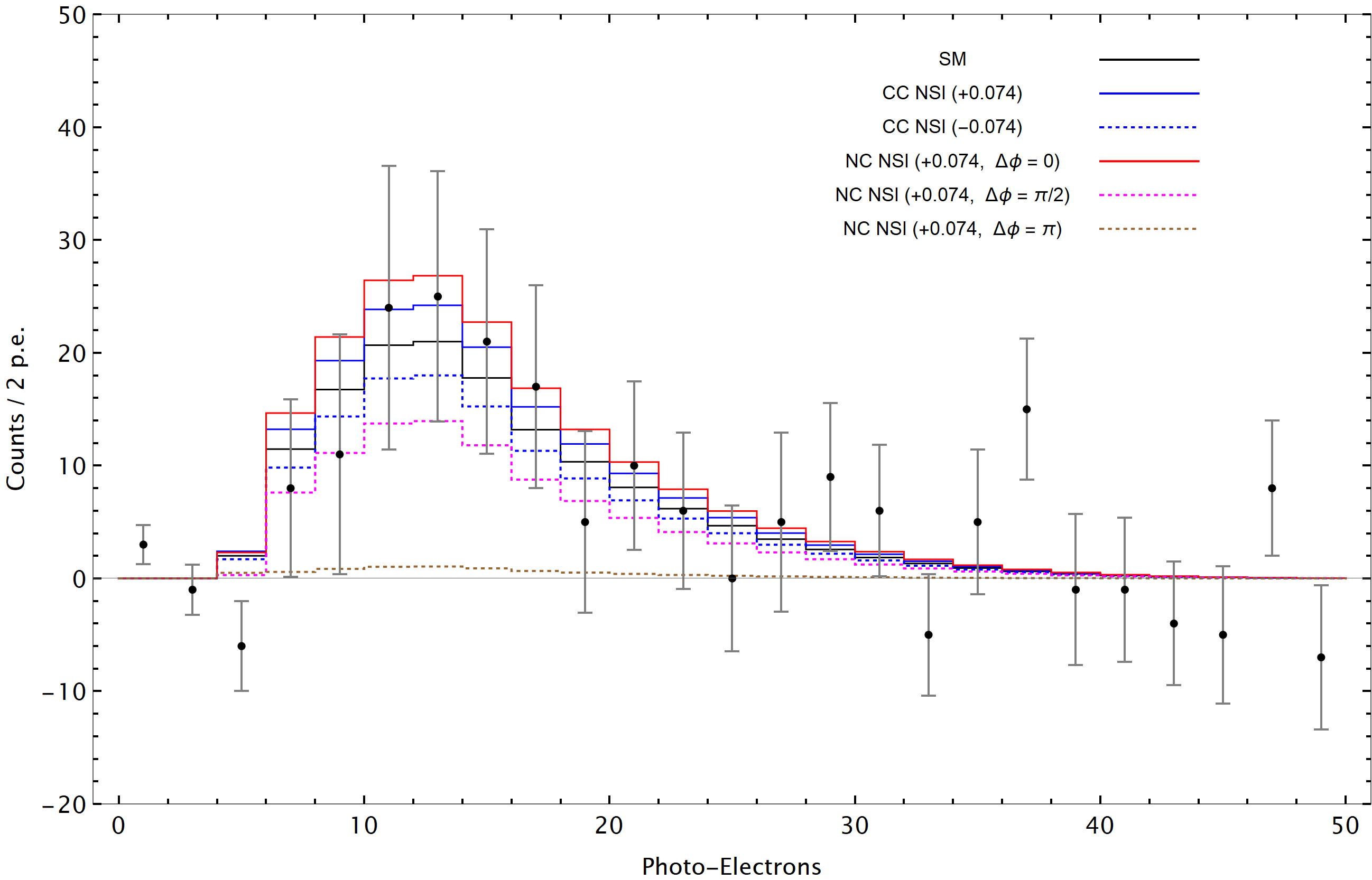}
\end{center}
\caption {Observed energy spectrum of COHERENT data in terms of photo-electrons together with the expected spectrum for SM, CC NSI and NC NSI with three choices of the new CP-phases. For the case CC NSI, the moduli were all taken $+0.074$ or $-0.074$ while setting the NC NSI to zero. For the case NC NSI all parameters were taken $+0.074$ with three choices for the CP-phases while setting all the CC NSI to zero. }
\label{fig:spectrum}
\end{figure}
\noindent
As explained before, due to the hermiticity of the NC Lagrangian in eqs.\ $(\ref{eq:NCLeff})$ and (\ref{eq:NCLeffVA}) the diagonal parameters $\varepsilon_{\beta \beta}^{qV}$ are real, while the flavor-changing parameters $\varepsilon_{\alpha \beta}^{qV}$ are complex and can be written in terms of  modulus and  phase as $|\varepsilon_{\alpha \beta}^{qV}| \,  \mathrm{e}^{i\phi_{\alpha\beta}^{qV}}$ for $\alpha \neq \beta$. After expanding the  terms, we can rewrite the weak charge in eq.\  $(\ref{eq:Qall})$ as
\begin{widetext}
\begin{eqnarray}
Q_{W\beta }^{2} &=& \left[Z(g_{p}^{V}+2\varepsilon _{\beta \beta
}^{uV}+\varepsilon _{\beta \beta }^{dV})+N(g_{n}^{V}+2\varepsilon _{\beta
\beta}^{dV}+\varepsilon _{\beta \beta }^{uV}) \right]^{2}  \nonumber \\
&&+\underset{\alpha \neq \beta }{\sum } \left[(2Z+N)^{2}
|\varepsilon_{\alpha\beta}^{uV}|^{2}+(Z+2N)^{2} |\varepsilon_{\alpha\beta}^{dV}|^{2}+2(2Z+N)(Z+2N)|\varepsilon_{\alpha\beta}^{uV}||\varepsilon_{\alpha\beta}^{dV}| \cos(\Delta\phi_{\alpha\beta})\right],
\end{eqnarray}%
\end{widetext}
where $\Delta\phi_{\alpha\beta}=
\phi_{\alpha\beta}^{uV}-\phi_{\alpha\beta}^{dV}$ is the relative phase of  $\varepsilon_{\alpha\beta}^{uV}$ and $\varepsilon_{\alpha \beta }^{dV}$. Notice that we have suppressed the superscripts ``$uV/dV$'' on the phases and ``$udV$'' on the relative phases. 
For $\nu_{\nu} / \nu_{\bar{\mu}}$ and $\nu_{e}$ respectively, $Q_{W\beta }^{2}$ is 
\begin{eqnarray}\label{eq:cp1}
Q_{W \mu/ \bar{\mu}}^{2} &=&\left[Z(g_{p}^{V}+2\varepsilon _{\mu \mu
}^{uV}+\varepsilon _{\mu \mu }^{dV})+N(g_{n}^{V}+2\varepsilon _{\mu \mu }^{dV}+\varepsilon _{\mu \mu }^{uV})\right]^{2}  \notag \\
&&+(2Z+N)^{2}
\left(|\varepsilon_{e \mu}^{uV}|^{2}+|\varepsilon_{\tau\mu}^{uV}|^{2}\right)+\left(Z+2N \right)^{2} (|\varepsilon_{e \mu}^{dV}|^{2}+|\varepsilon_{\tau\mu}^{dV}|^{2})   \\ \notag 
&&+2(2Z+N)(Z+2N) \left[|\varepsilon_{e \mu}^{uV}||\varepsilon_{e \mu}^{dV}|\cos(\Delta\phi_{e \mu})+|\varepsilon_{\tau \mu}^{uV}||\varepsilon_{\tau \mu}^{dV}|\cos(\Delta\phi_{\tau \mu})\right], \notag
\end{eqnarray}
\begin{eqnarray}\label{eq:cp2}
Q_{W e}^{2} &=&\left[Z(g_{p}^{V}+2\varepsilon _{ee
}^{uV}+\varepsilon _{ee}^{dV})+N(g_{n}^{V}+2\varepsilon _{ee }^{dV}+\varepsilon _{ee }^{uV})\right]^{2}  \notag \\
&&+(2Z+N)^{2}
(|\varepsilon_{e \mu}^{uV}|^{2}+|\varepsilon_{\tau e}^{uV}|^{2})+(Z+2N)^{2} (|\varepsilon_{e \mu}^{dV}|^{2}+|\varepsilon_{\tau e}^{dV}|^{2})  \\ &&+2(2Z+N)(Z+2N)\left[|\varepsilon_{e \mu}^{uV}||\varepsilon_{e \mu}^{dV}|\cos(\Delta\phi_{e \mu})+|\varepsilon_{\tau e}^{uV}||\varepsilon_{\tau e}^{dV}|\cos(\Delta\phi_{\tau e})\right].\notag
\end{eqnarray}%
Thus, in presence of NC NSI, the parameters to analyse are $|\varepsilon_{\mu\mu}^{u/dV}|, |\varepsilon_{ee}^{u/dV}|, |\varepsilon_{e \mu}^{u/dV}|,|\varepsilon_{\tau\mu}^{u/dV}|, |\varepsilon_{\tau e}^{u/dV}|, \Delta \phi_{e\mu}, \Delta\phi_{\tau\mu},\Delta\phi_{\tau e}$.

We can now take a look at the observable effects of the CC and NC NSI parameters including their CP-phases on COHERENT's energy spectrum. The result of this exercise is shown in fig.\ \ref{fig:spectrum}. The parameter values  are 
$\pm  0.074$ for the CC parameters given in eq.\ $(\ref{eq:fluxes})$ and $0.074$ for the modulus of the NC parameters in eqs.\  $(\ref{eq:cp1})$ and $(\ref{eq:cp2})$ with three choices of the relative CP-phases. 
As can be seen in eq.\ (\ref{eq:diff-crossecall}), the CP-terms are responsible for different interference effects in each case. When $\Delta \phi= 0$, there is  constructive interference, when $\Delta \phi= {\pi}$, there is destructive interference, while for $\Delta \phi= {\pi/2}$ the interference effects are zero.

One can expect that the constraints on the CC NSI parameters will be significantly worse than on the NC NSI. 
The main reason for this is that as soon as $\varepsilon^{uV}_{\alpha\beta}$ or $\varepsilon^{dV}_{\alpha\beta}$ are switched on, the proton number appears in the 
weak charge in eqs.\ (\ref{eq:fluxes}, \ref{eq:Qall}), which otherwise is very much suppressed due to the accidentally small $g_{p}^{V} \propto 1 - 4 \sin^2 \theta_W$. 
In contrast, CC NSI parameters appear as an overall ($1 + \varepsilon$) contribution to the flux, and hence there is less sensitivity to them. 
\section{Results and Discussion}\label{sec:results}
\noindent In this section, we will present the fits of the CC and NC parameters in the framework sat up so far. 
\subsection{Impact of CP-violating phases on the NC NSI parameter spaces}\label{sec:A}
To discuss the CP-effects more conveniently, we ignore first the flavor-diagonal terms and rewrite the cross section in terms of only the flavor-changing NSI parameters and their relative phases as 
\begin{widetext}
\begin{eqnarray}
\frac{d\sigma_{\beta}}{dT}(E_{\nu},T) &\simeq&\frac{G_{F}^{2}M}{\pi }[(Zg_{p}^{V}+Ng_{n}^{V})^{2}+\underset{\alpha \neq \beta }{\sum }[(2Z+N)^{2}|\varepsilon_{\alpha\beta}^{uV}|^{2}+(Z+2N)^{2}|\varepsilon_{\alpha\beta}^{dV}|^{2}   \\   &&+2(2Z+N)(Z+2N)|\varepsilon_{\alpha\beta}^{uV}||\varepsilon_{\alpha\beta}^{dV}|
 \cos(\Delta\phi_{\alpha\beta})]] \left( 1-\frac{MT}{2E_{\nu }^{2}}\right) F^{2}(q^{2})\,.\notag
\label{eq:diff-crossecall}
\end{eqnarray}
\end{widetext}
There are three relevant relative CP-phases, that is, $\Delta \phi_{e \mu}$, $\Delta \phi_{\tau \mu}$ and $\Delta \phi_{\tau e}$, occurring only in the flavor-changing terms. The phase $\Delta \phi_{e \mu}$ is related to  $\varepsilon_{e \mu}^{uV}$ and $\varepsilon_{e \mu}^{dV}$, and similarly $\Delta \phi_{\tau \mu}$ is related to $\varepsilon_{\tau \mu}^{uV}$ and $\varepsilon_{\tau \mu}^{dV}$ and $\Delta \phi_{\tau e}^{ud}$ to $\varepsilon_{\tau e}^{uV}$ and $\varepsilon_{\tau e}^{dV}$. \par
\begin{figure}[t]
\begin{center}
\includegraphics[width=7in, height=2.8in]{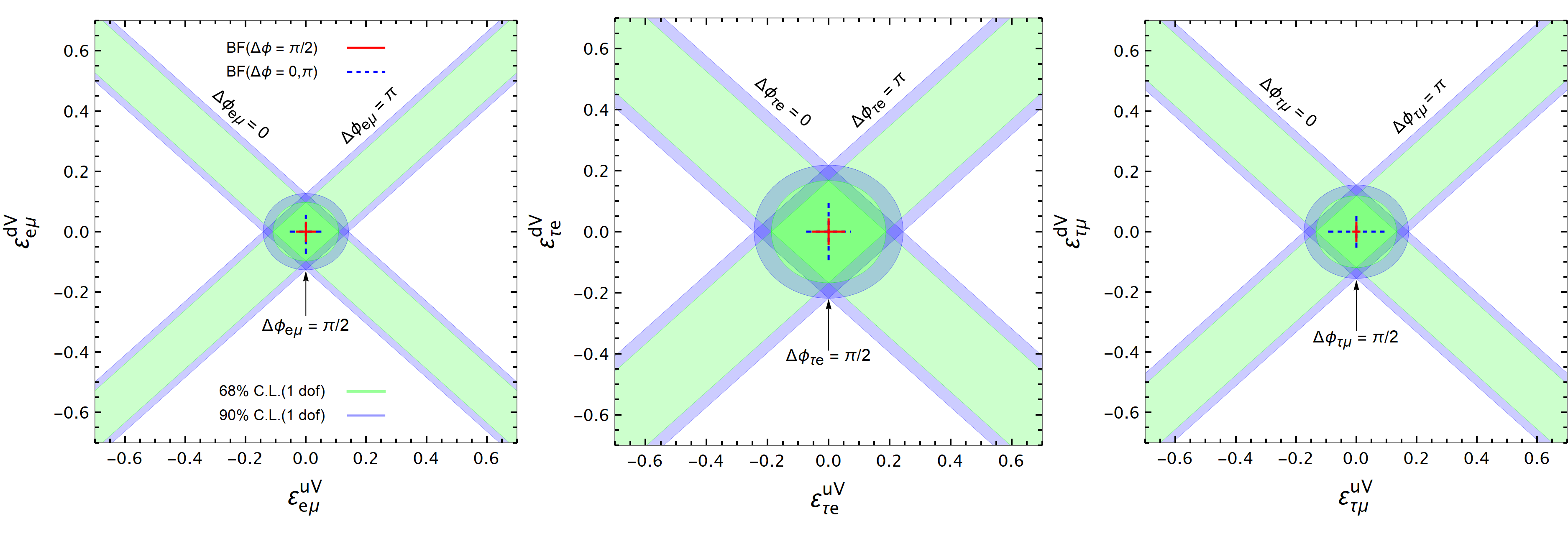}
\end{center}
\caption{\textbf{NC NSI:} 68\% and 90\% C.L.\ contour boundaries in the flavor changing NC NSI parameter spaces corresponding to  three possible sets of parameters with three extreme choices for the new CP-phases, that is, $\Delta\phi_{\alpha\beta}= \pi/2$ (the central elliptical contour), $\Delta\phi_{\alpha\beta}=0$ (the band with negative slope) and $\Delta\phi_{\alpha\beta}= \pi$ (the band with positive slope). The best-fit values are shown in red and blue colors. The relatively extended best-fits are due to the flat minimum in each case. The legend assignments in the left panel is applicable to all.}
\label{fig:Detector_2D_cp}
\end{figure}

For the fit we set one of the three $\varepsilon$ to zero and fit the other two 
for three extreme choices of the corresponding relative CP-phases, that is, $\Delta \phi= 0, {\pi}/2 \, \text{and} \, {\pi}$. The obtained results for the three parameter sets are shown in fig.\ \ref{fig:Detector_2D_cp}. In each case, the result for the choice corresponding to $\Delta {\phi}= 0$ was tacitly obtained before and reported in several previous papers, while the other two choices $\Delta {\phi}=\pi/2, \pi $ are presented for the first time in this work.\par 
In the case of no interference ($\Delta \phi = \pi/2$), the standard diagonal bands with both positive and negative slopes are transformed into the elliptical regions as can be seen for all three cases in fig.\ \ref{fig:Detector_2D_cp}. As a by-product of the no-interference choice, one can simultaneously constrain the two relevant absolute parameters in each case. As shown in blue and red, the lines at the center of all graphs corresponds to the degenerate minimum for each case.

\begin{figure}[t]
\begin{center}
\includegraphics[width=4.7in, height=4in]{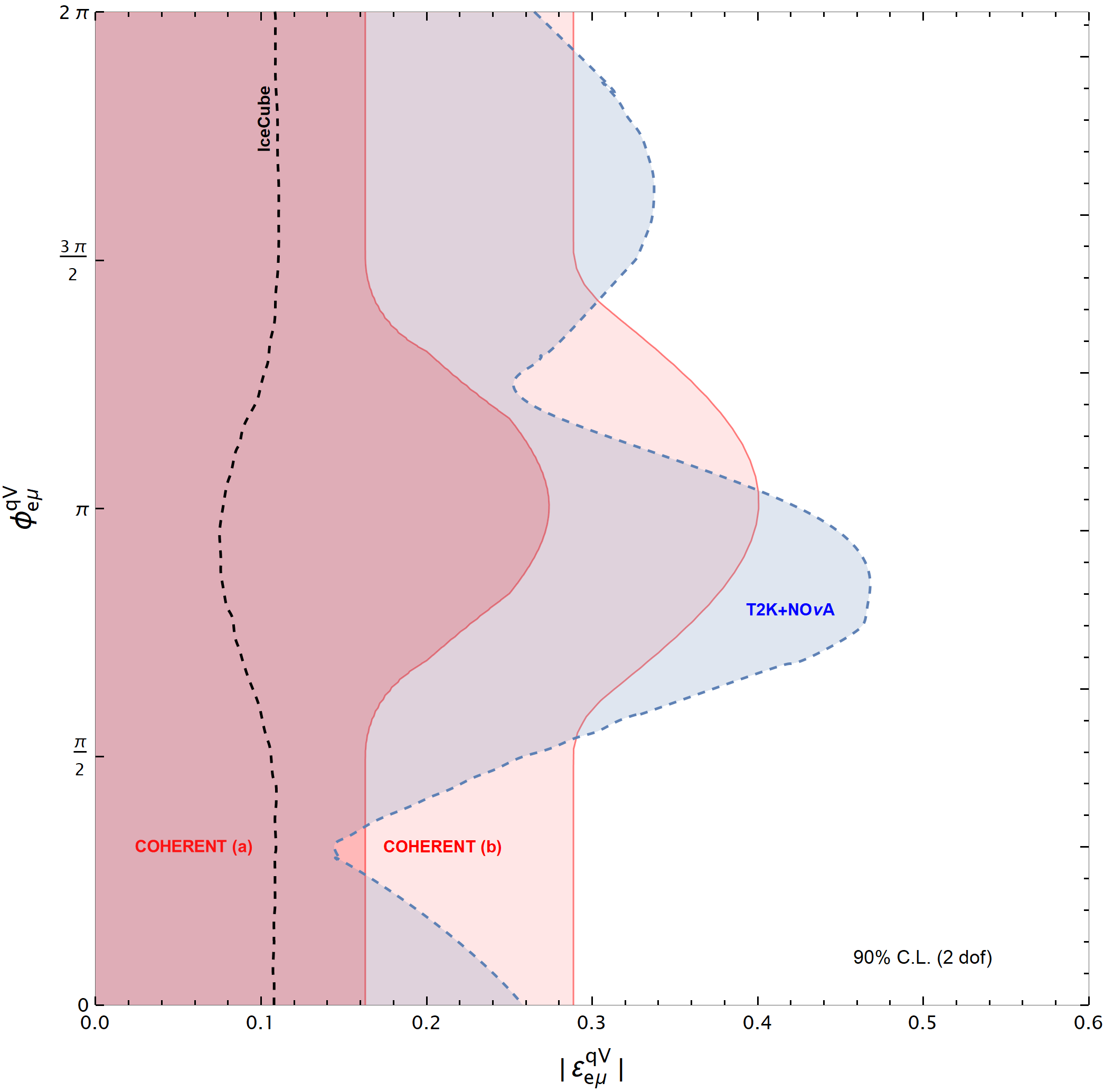}
\end{center}
\caption{\textbf{NC NSI:} 90\% C.L.\ contour boundaries in the parameter space of absolute NSI  parameter and the relevant CP-phase for the case when we set all NSI parameters equal zero except for one of the $\varepsilon_{e\mu}^{qV}$ ($q = u$ or $d$), over which we marginalize (COHERENT (a)) and for the case when we marginalize over all other parameters (COHERENT (b)). The overlaid curves for T2K+NO$\nu$A and IceCube were taken from refs.\  \cite{Denton:2020uda,ICslides} with normal ordering of the neutrino masses. For a realistic comparison, the T2K+NO$\nu$A and IceCube results of the absolute parameter boundaries on the horizontal axes were normalized for the two quark case. The region on the right side of all curves is the excluded region. }
\label{CP_vs_Abs}
\end{figure}

We continue by investigating the space of one particular set of parameters, namely the absolute value  $|\varepsilon_{e \mu}^{qV}|$ and the phase $ \phi_{e\mu}$, which are important for the long-baseline oscillation appearance and disappearance experiments. Very recently, there has been reported a $\sim 2\sigma$ discrepancy between T2K \cite{Abe:2011ks} and NO$\nu$A \cite{nova:exp} measurements of the standard 3$\nu$ oscillation CP-phase $(\delta)$ \cite{dunne:p,himmel:a}. In ref.\ \cite{Denton:2020uda}, it was argued that in the presence of NC NSI and a related new CP-phase this tension is reduced. We explore here the same parameter space relevant for the two long-baseline oscillation experiments. The result is shown in fig.\ \ref{CP_vs_Abs}, where we present the parameter range explaining the T2K/NO$\nu$A discrepancy, as well as an independent limit obtained by IceCube  \cite{ICslides}. Two fits of COHERENT data are performed by us. 

First, we take all other parameters equal to zero except one parameter over which we marginalize and fit the absolute parameter $|\varepsilon_{e \mu}^{qV}|$ and the corresponding phase $\phi_{e\mu}^{qV}$. This region is shown in dark red color and marked as "COHERENT (a)" in fig.\ \ref{CP_vs_Abs}. The marginalizing parameter is either $|\varepsilon_{e \mu}^{dV}|$ and its phase when we fit $|\varepsilon_{e \mu}^{uV}|$ and its phase, or $|\varepsilon_{e \mu}^{uV}|$ and its phase when we fit $|\varepsilon_{e \mu}^{dV}|$ and its phase. Second, we marginalize over all the other parameters and fit $|\varepsilon_{e \mu}^{uV}|$ and $\phi_{e\mu}^{uV}$ or $|\varepsilon_{e \mu}^{dV}|$ and $\phi_{e\mu}^{dV}$. This region is shown in light red color and marked as "COHERENT (b)" in fig.\ \ref{CP_vs_Abs}. This result is independent of the choice of the quark flavor due to the symmetry between terms for up and down quarks appearing in eq.\ (\ref{eq:diff-crossecall}).

As can be seen from fig.\ \ref{CP_vs_Abs}, marginalization mitigates the excluded region, while in the first case, the COHERENT data alone excludes a large parameters space allowed by NO$\nu$A and T2K, but relatively weaker than to IceCube. Even in case of COHERENT (b), COHERENT gives comparable or better constraints than NO$\nu$A and T2K in some parts of the parameters space. Also one can see from the figure, the parameter space of COHERENT for the first case (COHERENT (a)) shows similar behaviour to the IceCube. This points out how COHERENT is complementary to long-baseline experiments, and already tests part of the parameter space that explains the T2K/NO$\nu$A discrepancy. Note, however, that if there is only one $\varepsilon$, COHERENT has no sensitivity on any CP phase. 
\subsection{Constraints on CC NSI parameters from COHERENT data}\label{sec:B}
\begin{figure}[t]
\begin{center}
\includegraphics[width=7in, height=2.8in]{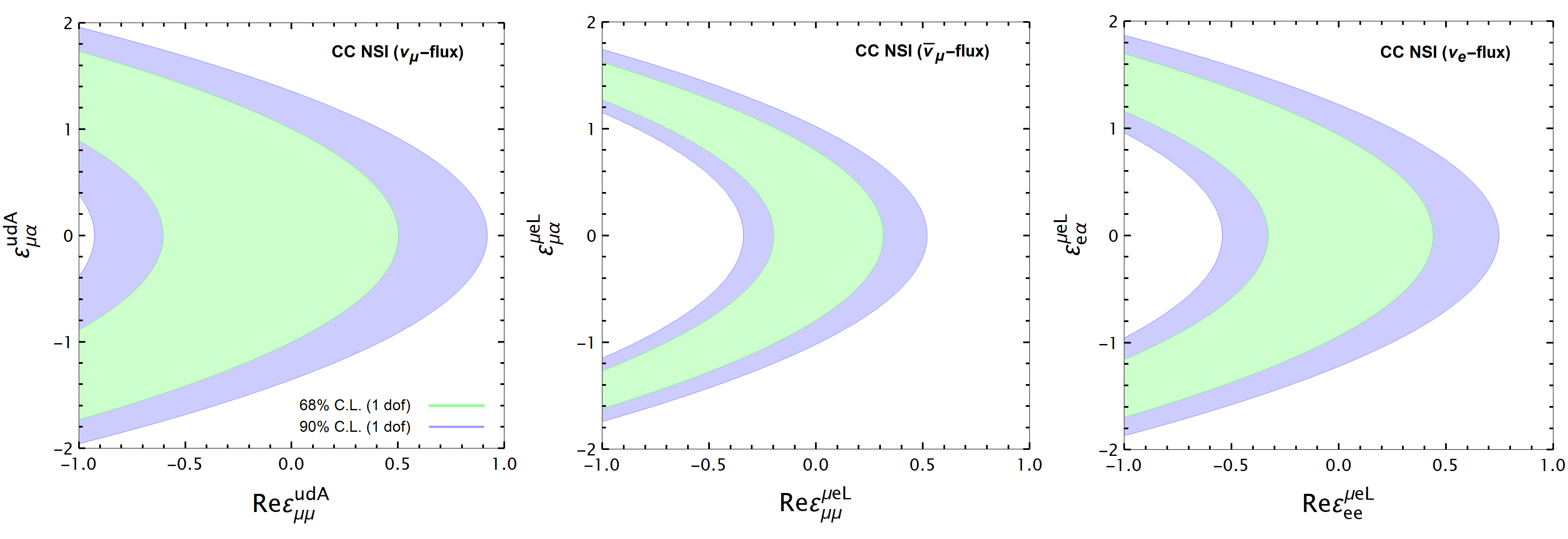}
\end{center}
\caption{Allowed regions of the CC NSI parameters relevant for the COHERENT setup considered in this work. In each figure, the index $\alpha$ of the label on $y$-axis corresponds to $e$, $\mu$ or $\tau$. Each figure corresponds to one of the three fluxes, $\nu_{\mu}\, (left)$, $\bar{\nu}_{\mu}\,(middle)$ and $\nu_{e}\,(right)$ as defined in eq.\ (\ref{eq:fluxes}).}
\label{fig:Source_2D}
\end{figure}

Now we use the COHERENT data to constrain the source CC NSI parameters related to  pion and muon decays. As can be seen in eq.\ (\ref{eq:fluxes}), each flux has two types of CC NSI parameters,  flavor conserving and flavor changing. Only the former interfere with the SM contribution. 
For each flux, we fit the real part of the flavor-conserving $\varepsilon$ and one 
flavor-changing  NSI parameter together, while setting  parameters in the other two fluxes to zero. The three fit results at $68\%$ and $90\%$ C.L.\ are shown in fig.\ \ref{fig:Source_2D}. The one parameter at-a-time constraints on each individual parameter are summarized in table \ref{tab1}. For comparison, we also give bounds from other studies, which were obtained from the kinematics of weak decays, CKM unitarity and branching ratios of meson decays. While the COHERENT constraints are weaker than those, we note that direct comparison with the other bounds from branching ratios and kinematics is not always straightforward, because those often involve charged leptons in contrast to neutrinos \cite{GonzalezGarcia:2001mp, Bergmann:1999pk}. 

Note that in eq.\ (\ref{eq:eventrt}) the real parts of the CC NSI parameters 
appear with a relative factor two compared to the squared absolute values, which explains the different scale on the axes in fig.\ \ref{fig:Source_2D}. 
Note further that the relative contribution to the total flux in COHERENT is 
50\% for $\bar{\nu}_{\mu}$, 31\% for ${\nu_e}$ and 19\% for ${\nu_{\mu}}$ \cite{Akimov:2017ade}. 
This reflects in the size of the constraints in the left $(\nu_{\mu})$, middle $(\bar{\nu}_{\mu})$ and the right $({\nu_e})$ panels of fig.\ \ref{fig:Source_2D}.
\begin{table*}[th]
\begin{center}
\begin{tabular}{c|c|c}
\hline \hline
parameter & COHERENT (this work) & other bounds  \\ \hline
$Re(\varepsilon _{\mu \mu }^{udA})$ & $[-0.9,0.9]$ &  $[-0.007, 0.012]\, (\textnormal{Br.})$
 \\ \hline
$\varepsilon _{\mu \alpha}^{udA}$& $[-1.3,1.3]
$ &$[-0.118, 0.118]\, (\textnormal{Br.})$  \\ \hline
$Re(\varepsilon _{\mu \mu }^{\mu eL})$ & $[-0.3,\ 0.5]$ & $[-0.030,\ 0.030]\, (\textnormal{Kin.})$
\\ \hline
$\varepsilon _{\mu \alpha}^{e\mu L}$ & $%
[-1.1,1.1]$ & $[-0.087,~0.087] \, (\textnormal{Osc.})$  \\
\hline
$Re(\varepsilon _{ee}^{\mu eL})$ & $[-0.5,~0.7]$ &$[-0.025,~0.025]\,  (\textnormal{Osc.})$
\\ \hline
$\varepsilon_{e\alpha }^{\mu eL}$ & $%
[-1.2,~1.2]$ &$[-0.030,~0.030]\, (\textnormal{Kin.})$ \\
\hline \hline
\end{tabular}%
\end{center}
\caption{One parameter at-a-time constraints at 90\% C.L.\ from this work for the CC NSI derived from fig.\ \ref{fig:Source_2D} and defined in eq.\ (\ref{eq:fluxes}) compared to  other studies \cite{Liu:2020emq, Biggio:2009nt}. The subscript $\alpha$ in the 1st column and 3rd, 5th, 7th row stands for $e, \mu, \tau$. In the column   ``other bounds'' the abbreviation ``Br.'' stands for branching ratios, ``Osc.'' stands for oscillations, ``Kin.'' stands for kinematics.}
\label{tab1}
\end{table*}
\subsection{Interplay between the CC NSI and the NC NSI at COHERENT and the LMA-Dark solution}
\begin{figure}[t]
\begin{center}
\includegraphics[width=6.2in, height=5.4in]{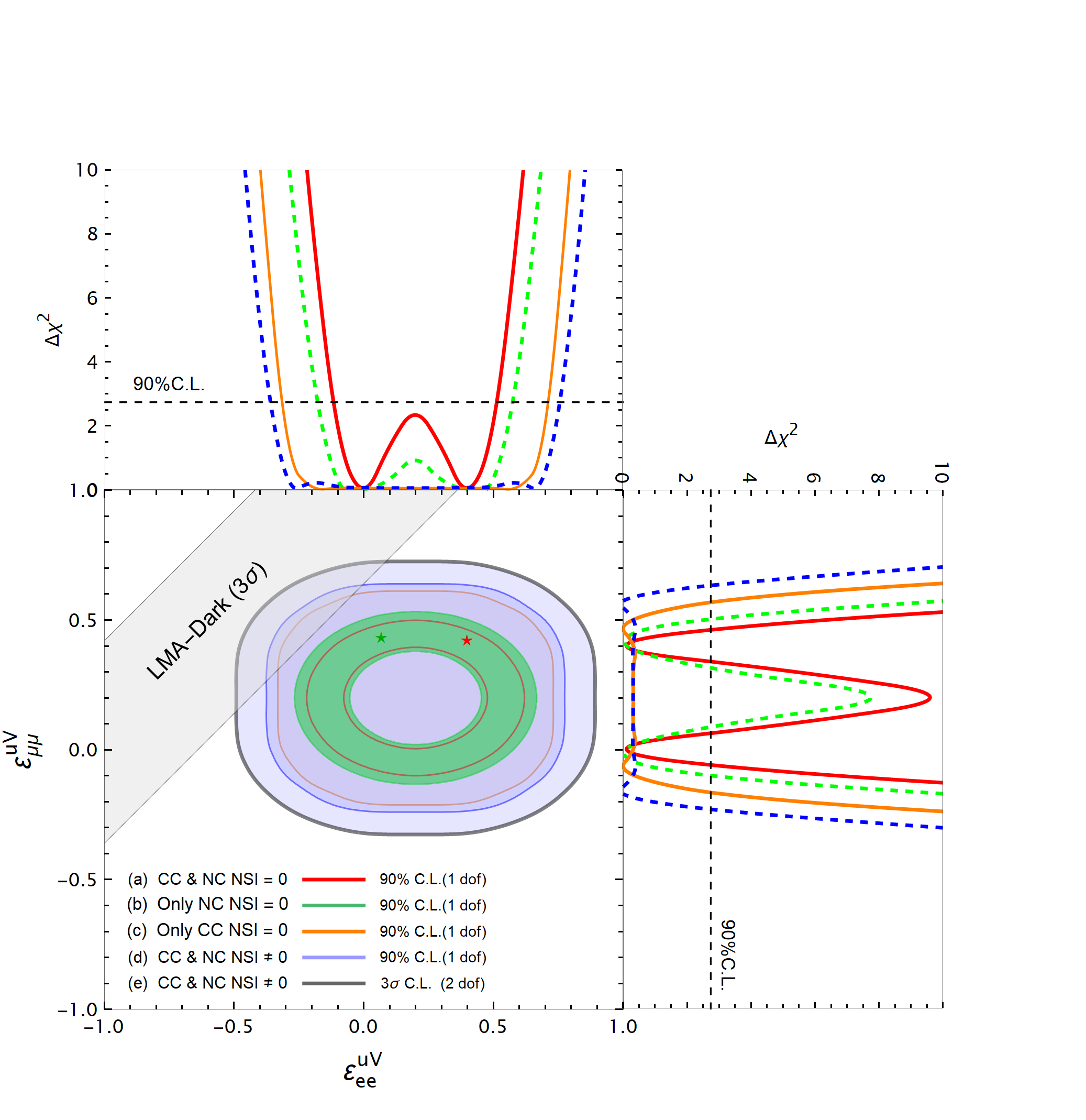}
\end{center}
\caption {\textbf{CC, NC NSI and CP-phases together:} 2-dimensional allowed regions for the flavor diagonal NC NSI parameters relevant for the LMA-Dark solution in solar data. For guidance of the best-fit values and the $90\%$ C.L.\ projections, we also provide one parameter at-a-time $\Delta\chi^2$ distribution for each fitting parameter in the top and right-side panels. Contour plots for case $(a)$ (red), $(b)$ (green), $(c)$ (orange), case $(d)$ (blue) were obtained at 90\% C.L.\ with $\Delta\chi^2$ for 1 dof while case  $(e)$ (black) was obtained at $3 \sigma$ for 2 dof C.L. The red and green stars corresponds to one of the two best-fit points for case $(a)$ and $(b)$, respectively. For case $(c)$, $(d)$ and $(e)$, the minima are flat as can also be seen in the one-dimensional plots. The legend colors for cases $(a)-(d)$ corresponds to both 2-dimensional and 1-dimension plots. See text for further details about the five cases and the fitting procedure. The $3\sigma$ diagonal band shows the LMA-Dark solution in solar data taken from ref. \cite{Coloma:2017egw}.}
\label{source&detector}
\end{figure}
For illustration on the interplay of CC and NC NSI parameters, we focus on fitting the two NC NSI parameters relevant for the LMA-Dark degeneracy existing in the solar oscillation data \cite{Miranda:2004nb}. This issue is related to the two possible solutions in the parameter space of the solar mixing parameters ($\theta_{12}$ and $\delta m_{21}^2$), where one solution is the standard $3 \nu$ mixing while the other one is caused by flavor-conserving NC NSI parameters during propagation and has, in particular $\theta_{12}>\pi/4$. The corresponding NSI parameters are $\varepsilon_{ee}^{uV}$ and $\varepsilon_{\mu \mu}^{uV}$, which are real. This possibility has been ruled out, in the pure effective operator limit in refs.\ \cite{Coloma:2017egw, Coloma:2017ncl, Denton:2018xmq, Liao:2017uzy,Khan:2019cvi, Giunti:2019xpr, Coloma:2019mbs, EstevesChaves:2021jct}. In the earlier papers \cite{Coloma:2017egw, Coloma:2017ncl, Denton:2018xmq, Liao:2017uzy,Khan:2019cvi,Giunti:2019xpr}, it was concluded that the LMA-Dark solution is excluded by the COHERENT data by at least $3\sigma$. Recently, ref.\ \cite{Coloma:2019mbs} presented a revised analysis and concluded that there is still room for the LMA-Dark solution which cannot be excluded by the CE$\nu$NS data. Very recently, ref.\ \cite{EstevesChaves:2021jct} has shown that  LMA-Dark is disfavored by $2.2\sigma$ in the presence of an extra phase for the corresponding flavor diagonal NSI parameters. \par 
In our following analysis, we will show how the significance level of the exclusion of the LMA-Dark solution gets affected in the presence of CC-NSI parameters and the new CP-phases. This is meant only as an illustration of the impact of a possible simultaneous presence of those. In principle one should fit the solar neutrino data in the presence of those parameters as well, which is beyond the scope of this work. From fig.\ \ref{source&detector}, one can see that after including the CC NSI and the CP-phases, the allowed boundaries extend towards the LMA-Dark region, which implies worsening of the exclusion significance of the LMA-Dark solution. A more concrete statement would require fitting solar and other oscillation data in combination with  coherent scattering data, which is beyond the scope of this work. 

Here we want to analyse the following aspects. First, we want to see the impact of the CC NSI parameters on the given flavor-conserving NC NSI in the fit. Second, we want to see effect of CP-phases on the given NC NSI parameters. Third, we want to see how the allowed region for the given parameters change with and without marginalization over all the other parameters. Fourth, how these three aspects change the significance level of excluding the LMA-Dark solution. We emphasize that we are not interested in fitting of all the NC NSI parameters in this study, which can be found in several other works, e.g.\ in refs.\  \cite{Giunti:2019xpr, Denton:2020hop}. Here we consider the following analysis as an example of how the above four motivations could be tested. To this aim we fit the two parameters $(\varepsilon_{ee}^{uV}$ and $\varepsilon_{\mu \mu}^{uV})$ with the following five choices:\newline
\textbf{(a)} Setting all the other NSI parameters equal to zero. \ \textbf{(b)} Marginalizing over all the real CC parameters in the range $(-0.1, 0.1)$ and absolute parameters in the range $(0.0, 0.1)$, while setting all the NC NSI parameters equal to zero.\ \textbf{(c)} Marginalizing over all real NC parameters in the range $(-0.1, 0.1)$ and absolute parameters in the range $(0.0, 0.1)$ with the three relative CP-phases in the range $(0, 2\pi)$ while setting all the CC NSI parameters equal to zero.\ \textbf{(d) \& (e)} Marginalizing over all  real parameters, both CC and NC NSI, in the range $(-0.1, 0.1)$ and absolute parameters, both CC and NC NSI, in the range $(-0.1, 0.1)$ with the relative CP-phases in the range  $(0, 2\pi)$.\par
The result of these fits is illustrated in fig.\ \ref{source&detector}. For each case mentioned above, we present our results of this analysis in two-dimensional allowed regions and in one-dimensional $\Delta\chi^2$ distributions in the top and right-side plots. The two-dimensional contour plots for the cases $(a)$-$(d)$ were obtained at $\Delta \chi^2 = 2.71$ $(90\%)$ for 1 dof in order to make a reasonable comparison with the 1-dimensional plots in the top and right-side panels while case $(e)$ was obtained with $\Delta \chi^2 = 11.83$ $(3\sigma)$ for 2 dof to compare with the corresponding $3\sigma$ LMA-Dark solution shown in fig.\ \ref{source&detector}. All the minima and the $90\%$ C.L.\ boundaries of the two-dimensional contours and the one-dimensional $\Delta\chi^2$ distributions in fig.\ \ref{source&detector} are consistent with each other. The best-fit points for cases $(a)$ and $(b)$ are shown with stars. As can be seen from the corresponding 1-dimensional plots, these two cases have absolute minima. For case $(c)$, $(d)$ and $(e)$, after including the CC, NC NSI and the CP-phases in the fit, the absolute minima are lost and we get a flat minimum. The range of the flat minimum for case $(c)$ and $(d)$ can be estimated from the projections of the 1-dimensional plots on the corresponding contour plots. Note that we have taken the same fitting procedure for the one-dimensional plots as for the two-dimensional plots in cases $(a)-(d)$ except one of the two parameters $(\varepsilon_{ee}^{uV}$, $\varepsilon_{\mu \mu}^{uV})$ was set to zero.\par
The effects of the CC NSI and CP-phases can be seen by comparing cases $(a)$ versus $(b)$, and $(c)$ versus $(d)\&(e)$ in fig.\ \ref{source&detector}. In each case when the CC NSI and CP-phases are included in the fits, the contour boundaries  broaden and extend towards the LMA-Dark solution. The CC NSI effects are seemingly small as compared to the NC NSI, but their effects are still there. 
As mentioned above for a fair comparison with the solar $3\sigma$ LMA-Dark solution, we also take the special case $(e)$ of the allowed region at $3\sigma$ C.L.\ $(\Delta \chi^2 = 11.83)$ (2 dof). We remind that $(e)$ corresponds to the case of including all parameters, that is, CC, NC and the CP-phases in the fit and thus is the most general case for testing the significance of the exclusion of the LMA-Dark solution.
\section{Summary and Conclusions}\label{sec:concl}
In  recent years, wide attention has been put to constrain new physics with CE$\nu$NS using  COHERENT data. There have also been several attempts to show how this process plays a complimentary role in resolving issues existing in oscillation  measurements of standard  mixing parameters which otherwise cannot be resolved by the oscillation experiments alone. Despite the important role of the observed process we find that two important aspects related to NSI, namely, the CC NSI at neutrino production and the new CP-violating phases associated with the NC NSI, are missing from  previous studies. A detailed analysis of these two aspects using the COHERENT data was the main goal of this paper. The procedure developed here for our fits of COHERENT data can of course be used for any future experimental setup. This paper focuses on the present situation. Detailed studies on future constraints will be presented elsewhere.

By including the CC NSI at the neutrino production and the CP-phases related to NC NSI at the detection, we have addressed two  issues in oscillation experiments, namely, the LMA-Dark solution and the tension between T2K and NO${\nu}$A measurements of the standard CP-phase ($\delta$). This is based on the fact that new CP-phases implied by NC NSI can be connected to measurements of the standard  CP-phase in running or future long-baseline neutrino oscillation experiments. This is another example on how scattering and oscillation experiments complement each other and can be used to resolve degeneracies. 
In addition, we have also constrained CC NSI.

As expected, the bounds on the CC NSI are not competitive with existing ones for reasons discussed in sections  \ref{sec:A} and \ref{sec:B}. However, future CE$\nu$NS experiments with larger precision and more statistics will certainly push the parameter space further, which will be an important independent test for the CC NSI models. On the other hand,  new CP-phases associated to NC NSI significantly change the limits on the absolute NC NSI parameter values and therefore need a careful treatment.\par
For the CP-effects, we have presented our results in terms of relative phases arising in the NC NSI interaction with the up and down quarks and in terms of terms of individual phases. For the first case, we analysed in detail how the allowed regions of the corresponding flavor-changing parameters are changed by including  relative CP-phases. In the second case, we chose one specific set of parameters, namely the absolute value and the associated individual CP-phase either for up or down quarks, which are relevant particularly for  T2K and NO${\nu}$A, but also for the IceCube. We performed analyses with and without including all other parameters in our fit to see their effects (see fig.\ \ref{CP_vs_Abs}) on the  oscillation measurements. In the one case (COHERENT (a)), COHERENT excludes a large parameter space which is allowed by NO$\nu$A and T2K while does relatively weaker with respect to the IceCube. Even in case of COHERENT (b), COHERENT shows competitive or better constraints than T2K and NO${\nu}$A.

To see the combined effects of all the CC, NC NSI and the associated CP-phases, we focused on two flavor-conserving parameters which are relevant for the solar oscillation data and which cause the LMA-Dark solution to the solar oscillation mixing parameters.\ We studied different cases as summarized in fig.\ \ref{source&detector}. If we include all the parameters in the fit, the previous $\gtrsim 3\sigma$ exclusion of the LMA-Dark solution is weakened and the allowed parameter space from COHERENT data extends almost to the center of the LMA-Dark solution. 

To conclude, CE$\nu$NS is not only a good way to probe the absolute NC NSI parameters, but also the CC NSI parameters  and the new CP-phases associated with the flavor-changing NC NSI parameters. Our analysis provides an independent method of testing those parameters and can contribute to resolve issues faced by the oscillation data.  

\begin{acknowledgments}
ANK was supported by Alexander von Humboldt Foundation under the postdoctoral fellowship program.
\end{acknowledgments}

\bibliography{biblio}

\end{document}